\documentclass[%
reprint,
superscriptaddress,
nofootinbib,
 amsmath,amssymb,
 aps,
]{revtex4-1}

\usepackage{graphicx}
\usepackage{dcolumn}
\usepackage{bm}

\usepackage{xcolor}
\usepackage[normalem]{ulem}

\usepackage[pagebackref=false]{hyperref}

\begin{document}

\title{Single-electron double quantum dots in bilayer graphene}

\author{L. Banszerus $^a$}
\thanks{These authors contributed equally to this work.}
\affiliation{JARA-FIT and 2nd Institute of Physics, RWTH Aachen University, 52074 Aachen, Germany,~EU}%
\affiliation{Peter Gr\"unberg Institute  (PGI-9), Forschungszentrum J\"ulich, 52425 J\"ulich,~Germany,~EU}
\author{S. M\"oller}
\thanks{These authors contributed equally to this work.}
\affiliation{JARA-FIT and 2nd Institute of Physics, RWTH Aachen University, 52074 Aachen, Germany,~EU}%
\author{E. Icking}
\thanks{These authors contributed equally to this work.}
\affiliation{JARA-FIT and 2nd Institute of Physics, RWTH Aachen University, 52074 Aachen, Germany,~EU}%
\affiliation{Peter Gr\"unberg Institute  (PGI-9), Forschungszentrum J\"ulich, 52425 J\"ulich,~Germany,~EU}

\author{K. Watanabe}
\author{T. Taniguchi}
\affiliation{ 
National Institute for Materials Science, 1-1 Namiki, Tsukuba, 305-0044, Japan 
}%

\author{C. Volk}
\affiliation{JARA-FIT and 2nd Institute of Physics, RWTH Aachen University, 52074 Aachen, Germany,~EU}
\author{C. Stampfer}
\affiliation{JARA-FIT and 2nd Institute of Physics, RWTH Aachen University, 52074 Aachen, Germany,~EU}%
\affiliation{Peter Gr\"unberg Institute  (PGI-9), Forschungszentrum J\"ulich, 52425 J\"ulich,~Germany,~EU}%

\date{\today}


\begin{abstract} 
We present transport measurements through an electrostatically defined  bilayer graphene double quantum dot in the single electron regime. With the help of a back gate, two split gates and two finger gates we are able to control the number of charge carriers on two gate-defined quantum dot independently between zero and five. 
The high tunability of the device meets requirements to make such a device a suitable building block for spin-qubits. In the single electron regime, we determine interdot tunnel rates on the order of 2~GHz. Both, the interdot tunnel coupling, as well as the capacitive interdot coupling increase with dot occupation, leading to the transition to a single quantum dot. Finite bias magneto-spectroscopy measurements allow to resolve the excited state spectra of the first electrons in the double quantum dot; being in agreement with spin and valley conserving interdot tunneling processes.
\end{abstract}

\maketitle
\def\thefootnote{a}\footnotetext{Corresponding author: luca.banszerus@rwth-aachen.de}\def\thefootnote{\arabic{footnote}}

Electrostatically defined quantum dots (QDs) offer a compelling platform for spin-qubit-based quantum computation~\cite{Loss1998Jan}. For that purpose, QDs in semiconductor heterostructures mainly based on GaAs~\cite{Nowack2011Sep,Bluhm2010Dec} and silicon~\cite{Veldhorst2015Oct,Vandersypen2017Sep} have been studied intensively. For example, high-fidelity single-qubit~\cite{Yoneda2017Dec} and two-qubit~\cite{Watson2018Feb,Zajac2018Jan,Huang2019Apr} gate operations have been recently demonstrated for silicon qubit devices.
Graphene has been early identified as an alternative attractive material platform for spin-qubits thanks to its low nuclear spin densities, weak hyperfine coupling and weak spin–orbit interaction promising long spin decoherence times~\cite{Trauzettel2007Feb}.
Physically etched graphene quantum devices including quantum dots~\cite{Ihn2010Mar,Guttinger2010Sep} and double quantum dots (DQDs)~\cite{Molitor2010Mar,Volk2011Aug} have been studied for about a decade. 
Major achievements include the implementation of charge detection~\cite{Guttinger2008Nov,Wang2010Dec}, the observation of spin-states~\cite{Guttinger2010Sep} and the measurement of charge relaxation times~\cite{Volk2013Apr}.
However, the influence of disorder, in particular edge disorder~\cite{Engels2013Aug,Bischoff2012Nov}, prevented a precise control of the number of charge carriers on individual QDs making spin-qubit implementation impossible. 

The advancements in ultra-clean van der Waals heterostructures and in particular the use of local graphite gates allowed for the development of electrostatically defined bilayer graphene (BLG) quantum point contacts~\cite{Overweg2018Jan,Kraft2018Dec,Lee2019Nov,Banszerus2019Nov}, quantum dots~\cite{Eich2018Jul,Kurzmann2019Aug,Kurzmann2019Jul} and double quantum dots (DQDs)~\cite{Banszerus2018Aug,Eich2018Aug}. 
While single-electron and hole occupation has been demonstrated recently for individual QDs~\cite{Eich2018Jul}, the number of charge carriers in DQDs could not be controlled yet~\cite{Banszerus2018Aug,Eich2018Aug}. The precise control of the number of charge carriers is, however, a requirement for qubit operations in a semiconductor QD device.

Here, we show single electron occupation of a bilayer graphene DQD. The electrostatically defined DQD allows for a high tunability of the electrochemical potential such that we can precisely control the number of electrons on each of the QDs independently down to zero. The gate voltages tune the interdot tunnel coupling such that a gradual transition from the DQD into a larger single QD is achieved. Furthermore, we can shape the potential landscape to form an ambipolar n-p-n triple QD. By finite bias magneto-spectroscopy measurements we resolve the excited state spectrum of the DQD. The absence of excited state transitions at the $(0,0) \rightarrow (1,1)$ triple point is in agreement with spin and valley conserving interdot tunneling processes.

\begin{figure*}[]
	\centering
\includegraphics[draft=false,keepaspectratio=true,clip,width=0.99\linewidth]{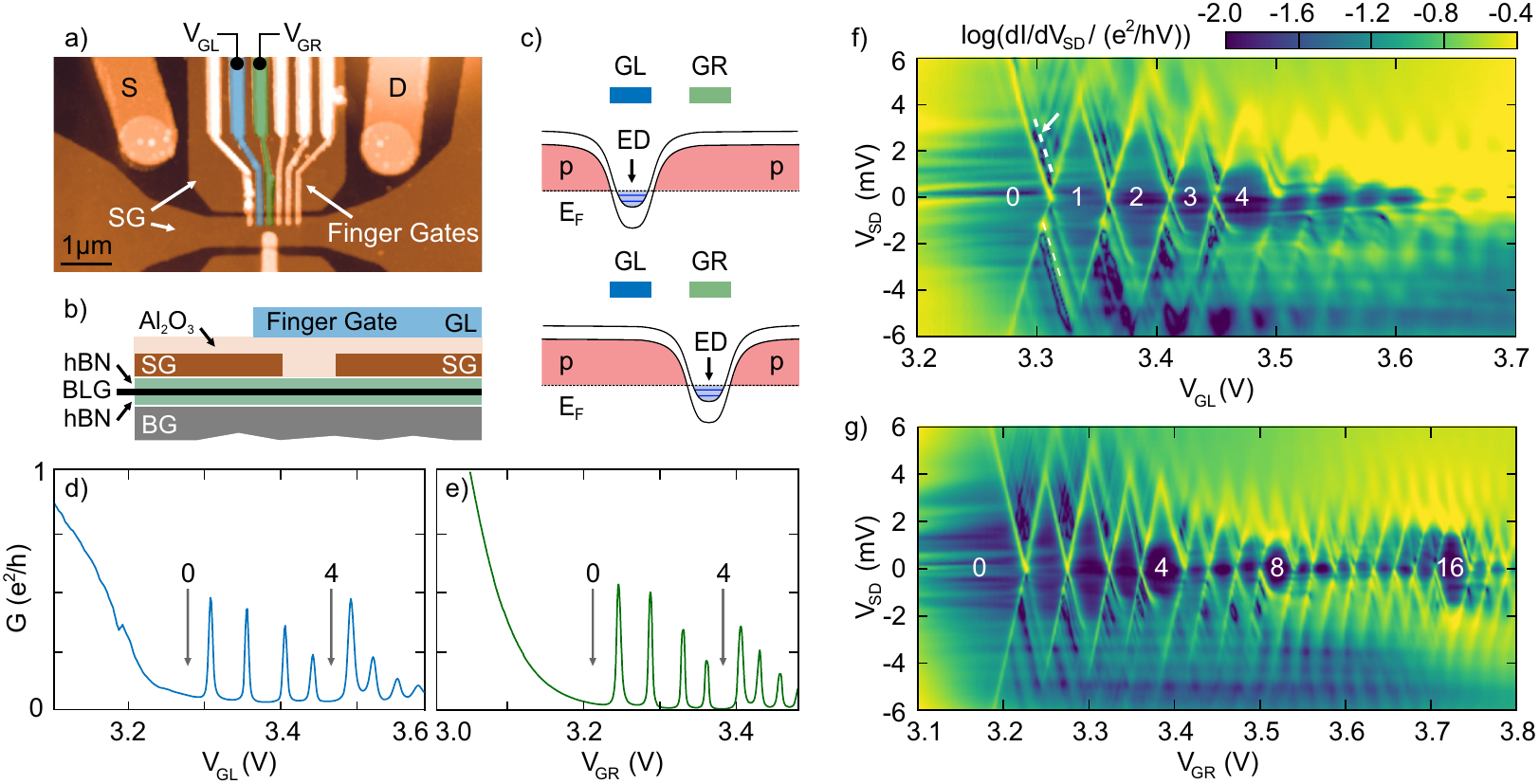}
\caption[Fig01]{  
\textbf{(a)}  Atomic force microscope image, showing the gate layout of the device. The source (S) and drain (D) contacts are connected to the bilayer graphene through etched vias in the hBN. The gate stack contains split gates (SG) with a separation of 50~nm and on top, separated by Al$_2$O$_3$, six parallel finger gates with a gate separation of 50~nm and a width of 100~nm. The gates GL and GR (color coded) are used to control the QDs discussed in this work.
\textbf{(b)} Schematic cross section of the device. The heterostructure is stacked in the sequence: graphite back gate, hBN, BLG, hBN. Two layers of metal gates are fabricated on top.
\textbf{(c)} Schematics of the band profile along the one-dimensional channel illustrating the formation of QDs. Applying a positive voltage to finger gate GL (GR), a p-n-p junction is formed, where a n-doped island forms a QD under GL (GR) separated from p-type reservoirs by two p-n junctions acting as tunneling barriers.
\textbf{(d)} Conductance through a QD formed under GL as a function of the gate voltage $V_\mathrm{GL}$. $V_\mathrm{SD} = 1$~mV.
\textbf{(e)} Measurement as in panel (d) for the QD formed under GR. Numbers indicate the electron occupation of the QDs.
\textbf{(f,g)} Finite bias spectroscopy measurements for both QDs formed under GL and GR. Numbers indicate the electron occupation in the regions of Coulomb blockade.}
\label{f1}
\end{figure*}

The studied device consists of a BLG flake encapsulated in two hexagonal boron nitride (hBN) crystals, fabricated by mechanical exfoliation and a dry van-der-Waals pick-up technique~\cite{Engels2014Sep,Wang2013Nov}. The heterostructure is placed on a graphite flake, acting as a back gate~\cite{Banszerus2018Aug}. On top of the stack, Cr/Au split gates are used to define a one-dimensional (1D) channel with an approximate width of 50~nm between the source and drain contacts. Separated from the split gates by a 30~nm thick layer of atomic layer deposited Al$_2$O$_3$, we fabricate 100~nm wide Cr/Au finger gates, separated by around 50~nm to define individual quantum dots. Fig.~\ref{f1}(a) shows an atomic force micrograph of the device highlighting the gate structure and the contacts. A schematic cross section of the device is shown in Fig.~\ref{f1}(b). For details of the fabrication process see Ref.~\cite{Banszerus2018Aug}. 
All measurements are performed in a $^{\text{3}}$He/$^{\text{4}}$He dilution refrigerator at a base temperature of around 10~mK using a combination of DC-measurements and standard low-frequency lock-in techniques.

We use the back gate and the split gates to apply an out-of-plane electrostatic displacement field in order to open a band gap in the BLG below the split gate area~\cite{Oos2007Dec}. At a back gate voltage $V_\mathrm{BG}=-1.7~V$ and a split gate voltage $V_\mathrm{SG}=1.84~V$ the Fermi energy lies within the band gap underneath the split gates leaving a p-type 1D channel in between them. 
As described in earlier work~\cite{Banszerus2018Aug,Eich2018Jul}, we can use the finger gate voltages ($V_\mathrm{GL}$ and $V_\mathrm{GR}$) to form a p-n-p band profile along the channel where a n-type QD is surrounded by p-type reservoirs, separated by the band gap acting as tunneling barriers (see e.g. the illustration Fig.~\ref{f1}(c)).

Figs.~\ref{f1}(d) and~\ref{f1}(e) show the conductance through the channel defined by the split gates as a function of the finger gate voltages $V_\mathrm{GL}$ and $V_\mathrm{GR}$, respectively (at a bias voltage of $V_\mathrm{SD}=1~$mV). The obtained conductance traces are qualitatively very similar. A series of Coulomb peaks proofs the formation of a QD below each of the finger gates and the sequence is in agreement with a fourfold shell filling due to spin and valley degeneracy in BLG~\cite{Eich2018Jul}. With decreasing finger gate voltage, first the QD is fully depleted, then the conductance increases without showing additional Coulomb peaks. This can be explained by pushing the Fermi level into the valence band, thus forming an unipolar p-type region along the entire channel between the split gates. 

Finite bias spectroscopy measurements of these QDs are shown in Figs.~\ref{f1}(f) and \ref{f1}(g). From these data we can extract addition energies of $E_\mathrm{add,L} \approx 4.3$~mV and $E_\mathrm{add,R} \approx 4.1$~mV from the first to the second electron in the left and the right QD, respectively. The extracted gate lever arms are $\alpha_L = 0.08$ and $\alpha_R = 0.09$ corresponding to gate capacitances of $C_\mathrm{GL} = 3$~aF and $C_\mathrm{GR} = 3.5$~aF. Thus, both QDs have similar charging energies and similar lever arms, indicating a rather high device uniformity and similar geometric sizes of the two QDs. In a simplified approximation, we consider a parallel plate capacitor formed by the QD and the metal finger gate controlling the dot, separated by the dielectric layers of $t_{\mathrm{hBN}} \approx 20$~nm of hBN and $t_{\mathrm{Al}_2\mathrm{O}_3} \approx 30$~nm of Al$_2$O$_3$. The finger gate capacitance is therefore approximated by $C_\mathrm{FG} \approx \epsilon_0 d^2/4 (t_{\mathrm{hBN}}/\epsilon_{\mathrm{hBN}}+t_{\mathrm{Al}_2\mathrm{O}_3}/\epsilon_{\mathrm{Al}_2\mathrm{O}_3})$, where $\epsilon_{\mathrm{hBN}}=4$ and $\epsilon_{\mathrm{Al}_2\mathrm{O}_3}=9$ are the dielectric constants and $d$ is the QD diameter. This results in QD diameters of around $d \approx$~60~nm and 65~nm, respectively, which is in reasonable agreement with the lithographic device dimensions.

The variation in size of the Coulomb diamonds resembles a fourfold shell-filling due to spin and valley degeneracy. Outside the regions of Coulomb blockade, we resolve a spectrum of excited states. 
Following Refs.~\cite{Banszerus2018Aug,Volk2011Aug} we can estimate the orbital excited state energies by $\Delta = 2\hbar^2/d^2 m^* \approx 1.1-1.3$~meV ($m^* = 0.033~m_e$ is the effective electron mass in BLG) from the QD diameter $d$. This energy is in agreement with the prominent excited state indicated by the white arrow in Fig.~\ref{f1}(f).

\begin{figure*}[]
	\centering
\includegraphics[draft=false,keepaspectratio=true,clip,width=0.95\linewidth]{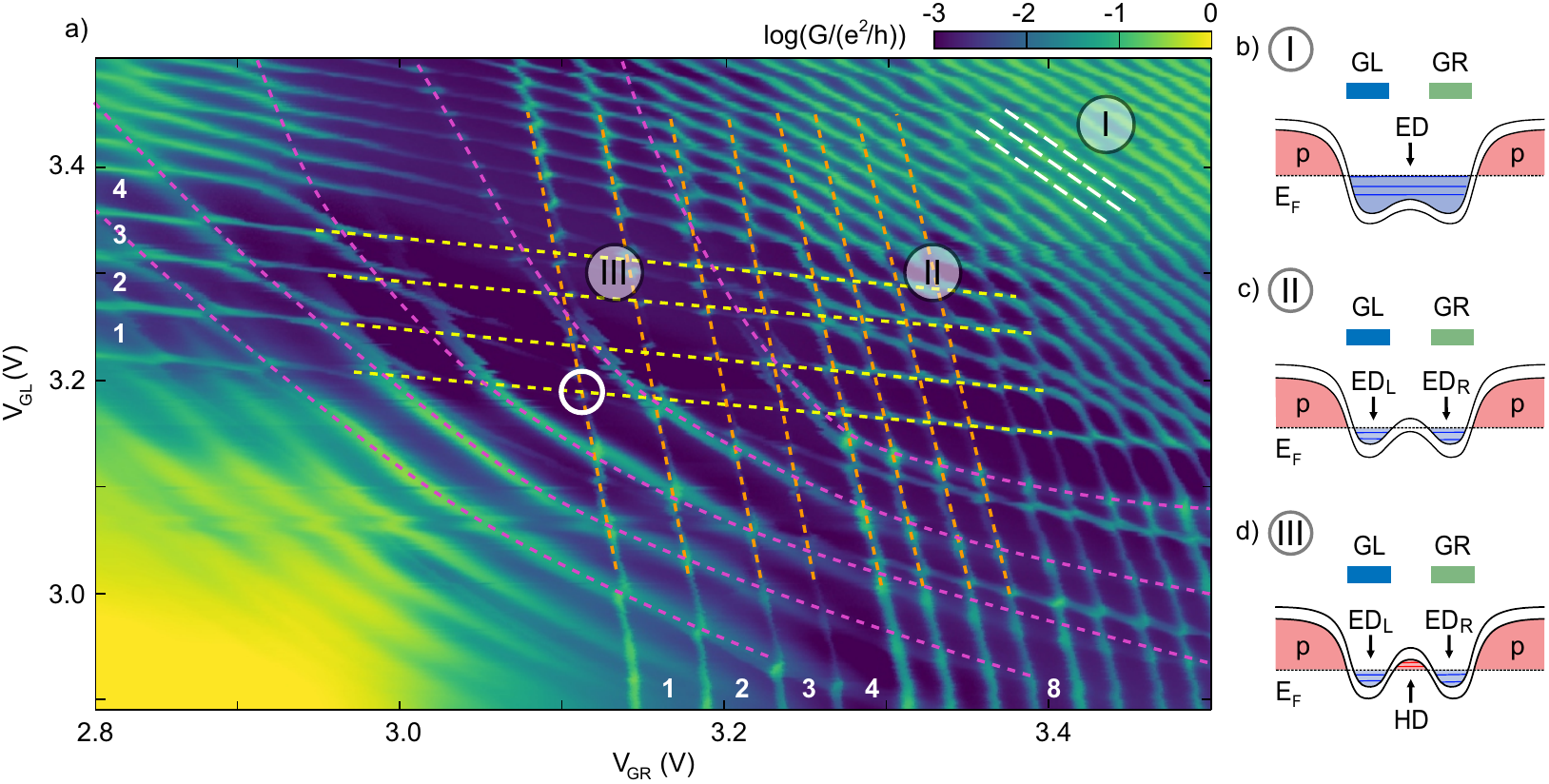}
\caption[Fig02]{
\textbf{(a)} Charge stability diagram showing the conductance through the channel as a function of the finger gate voltages $V_\mathrm{GL}$ and $V_\mathrm{GR}$ at a constant bias voltage of $V_\mathrm{SD} = 0.5$~mV, back gate voltage $V_\mathrm{BG}$~=~-1.7~V and split gate voltage  $V_\mathrm{SG}$~=~1.84~V. Dashed lines are a guide to the eye highlighting charge transitions of electron and hole QDs.
\textbf{(b-d)} Schematic illustrations of the band profile along the 1D channel defined by the split gates for different finger gate voltages.
\textbf{(b)} At elevated finger gate voltages $V_\mathrm{GL}$, $V_\mathrm{GR}$ one elongated QD is formed (see regime I in panel (a)).
\textbf{(c)} Band profile of a DQD: Using the finger gates GL and GR, a p-n-n-p junction is formed, where two n-doped islands form a n-n DQD surrounded by p-type reservoirs. Between the two QDs a tunnel barrier is formed as the Fermi level crosses the band gap (see regime II in panel (a)).
\textbf{(d)} Band profile of an ambipolar triple dot: At lower $V_\mathrm{GL}$, $V_\mathrm{GR}$ a p-type island is formed between the two n-type QDs resulting in a n-p-n triple dot (see regime III in panel (a)).
} 
\label{f2}
\end{figure*}

Fig.~\ref{f2}(a) shows the current through the channel as a function of the finger gate voltages $V_\mathrm{GL}$ and $V_\mathrm{GR}$ at fixed $V_\mathrm{BG}$ and $V_\mathrm{SG}$.
At high gate voltages $V_\mathrm{GL} \approx V_\mathrm{GR} > 3.4~$V a single QD is formed as illustrated by the band diagram in Fig.~\ref{f2}(b) (see regime I in Fig.~\ref{f2}(a)). The Fermi energy crosses the band gap between the DQD and the p-type reservoirs. The resulting p-n-junctions serve as tunnel barriers. In this regime, only one type of transition with a slope of approximately~-0.9 (see white dashed lines in Fig.~2(a)) can be observed in the charge stability diagram.
Due to cross capacitance effects, the gate voltages do not only influence the potential landscape directly underneath them but also the region in between. By reducing the finger gate voltages $V_\mathrm{GL}$ and $V_\mathrm{GR}$, the region between the left and right gate is tuned into the band gap, resulting in a DQD by breaking apart the larger QD. 
Around $V_\mathrm{GL} \approx V_\mathrm{GR} \approx 3.3$~V the typical signature of a charge stability diagram of a DQD can be observed. Two n-type QDs are formed under the finger gate GL and GR as indicated in Fig.~\ref{f2}(c). The almost horizontal and vertical charge addition lines (see yellow and orange dashed lines in Fig.~\ref{f2}(a)) correspond to charge transitions in each of the QDs. The low values of the relative gate lever arms of 0.12 and 0.15 indicate a high symmetry of the device and a rather small capacitive cross-talk (see regime II in  Fig.~\ref{f2}(a)). 

Lowering the gate voltages further depletes the DQD. Below $V_\mathrm{GL} \approx 3.2$~V and $V_\mathrm{GR} \approx 3.15$~V no further horizontal or parallel charge addition lines are present indicating both QDs have been emptied. 
As the current increases, we can exclude the possible effect that the lowered gate voltages render the tunnel barriers opaque suppressing the current below a detectable limit and thus masking the charge transitions. 

Additionally, a low finger gate voltage lifts the valence band above the Fermi level in the region between the gates ${\mathrm{GL}}$ and ${\mathrm{GR}}$, such that a hole QD (HD) is formed between the two QDs as illustrated in Fig.~\ref{f2}(d) (see regime III in Fig.~\ref{f2}(a)). Charge addition lines of this QD show up as curved lines in the charge stability diagram. The gate voltage dependent lever arm indicates that the hole QD is moved along the channel as a function of $V_\mathrm{GL}$ and $V_\mathrm{GR}$. The central region of Fig.~\ref{f2}(a) is in agreement with the charge stability diagram of a n-p-n triple QD.
In the low voltage regime ($V_{\mathrm{GL}} \lesssim 3$~V and $V_{\mathrm{GR}} \lesssim 2.9$~V), the device undergoes the transition from a hole QD to a homogeneous p-type channel as the tunnel coupling to the hole QD to the p-type leads increases significantly with lowered gate voltages.

\begin{figure}[]
	\centering
\includegraphics[draft=false,keepaspectratio=true,clip,width=0.95\linewidth]{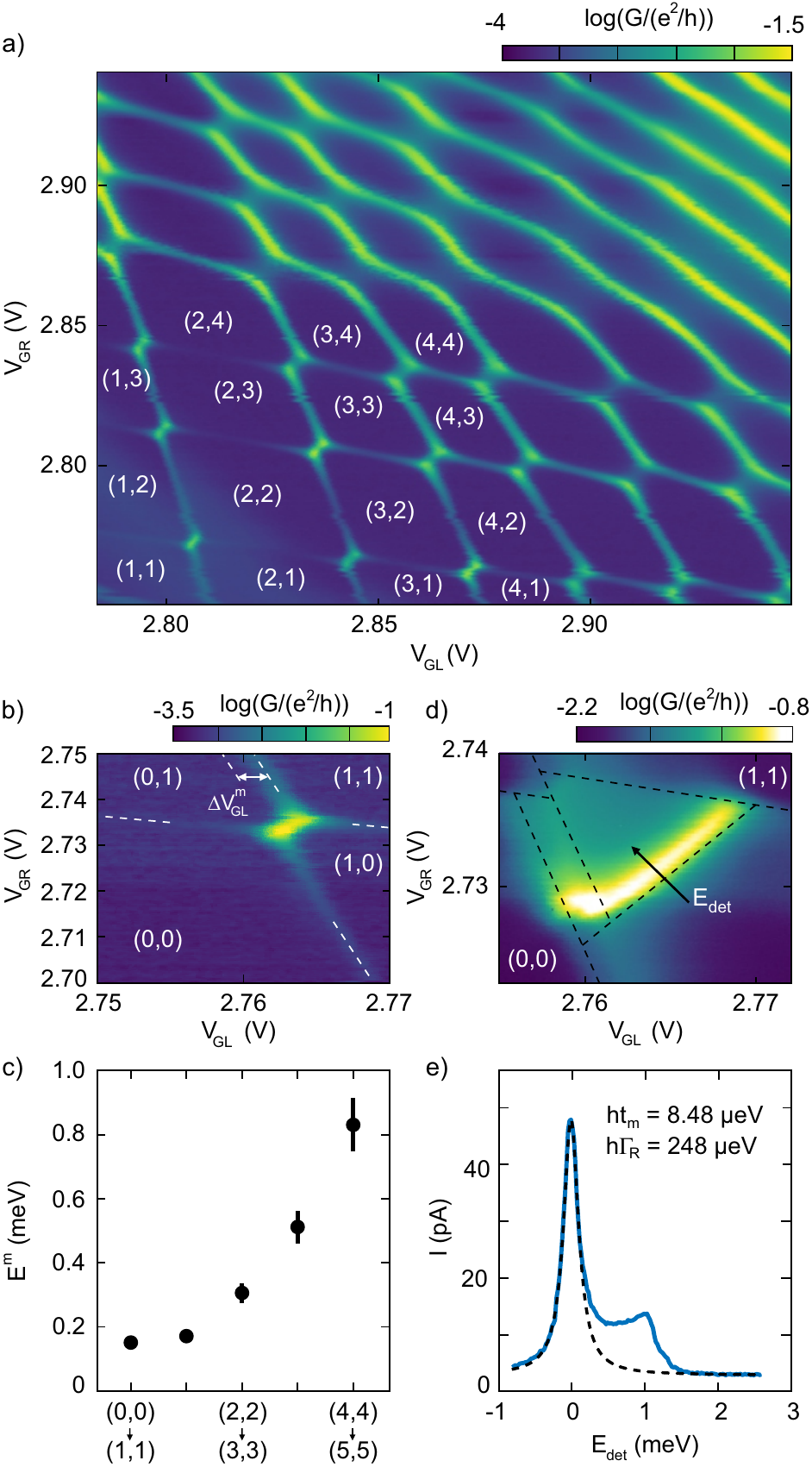}
\caption[Fig03]{
\textbf{(a)} Charge stability diagram of the DQD in the few electron regime measured at $V_\mathrm{SD} = 0.2$~mV, $V_\mathrm{BG}$~=~-1.5~V and $V_\mathrm{SG}$~=~1.62~V. ($N$, $M$) indicate the number of electrons on the left and right QD respectively.
\textbf{(b)} Charge stability diagram of the DQD in the single-electron regime ($V_\mathrm{SD} = 0.1$~mV).
\textbf{(c)} Capacitive interdot coupling of the DQD as function of the electron occupation for transitions with symmetric dot occupations (see panels (a) and (b)).
\textbf{(d)} Charge stability diagram as in panel (b) but at elevated bias ($V_\mathrm{SD} = 1$~mV).
\textbf{(e)} Current through the DQD (solid line) measured along the interdot detuning axis crossing the (0,1)-(1,0) charge transition as indicated by the arrow in panel (d). The peak at zero detuning has been fitted (dashed line) according to a model following Refs.~\cite{Fringes2012Jan,Liu2010May,Wiel2002Dec,Stoof1996Jan} in the limit of $eV_\mathrm{SD} >\!\!> k_\mathrm{B}T$ and $t_\mathrm{m} <\!\!< \Gamma$.
} 
\label{f3}
\end{figure}

Fig.~\ref{f3}(a) shows a high resolution charge stability diagram of the device in the DQD regime recorded at a low bias voltage ($V_\mathrm{SD} = 0.2$~mV). Careful tuning of the back gate and the split gate voltages 
allows to shift away the hole QD transitions, resulting in true DQD. The charge stability diagram shows the signature of a clean DQD in the low electron regime (a plot similar to Fig.~2(a) for this back gate and the split gate voltage is shown in the Supplementary
Fig. S1). The number of electrons can be controlled from single electron occupation up to about five electrons per QD. The device undergoes a transition to a single QD with increasing gate voltages. 

From the charge stability diagram and finite bias spectroscopy data we can extract gate lever arms of $\alpha_L = 0.11$ and $\alpha_R = 0.1$ and addition energies of $E_\mathrm{add,L} = 3.9$~meV and $E_\mathrm{add,R} = 3.8$~meV (see Supplementary Fig.~S2(a))~\cite{Molitor2009Jun,Volk2011Aug,Banszerus2018Aug}. 
Following the model of a parallel plate capacitor, these values correspond to QD diameters of 73~nm and 71~nm which is slightly larger compared to the single QDs formed individually, where we determined 60 and 65~nm. 
Furthermore, the charge stability diagram allows to determine the mutual capacitive coupling $E_\mathrm{m}$ of the DQD which is depicted schematically in Fig.~\ref{f3}(b).
Fig.~\ref{f3}(c) shows $E_\mathrm{m}$ for different pairs of triple points with symmetric electron occupation, i.e. transitions $(N,N) \rightarrow (N+1,N+1)$. The capacitive coupling increases significantly with the number of charge carriers and thus the applied gate voltages. This is in contrast to earlier experiments on etched single-layer~\cite{Molitor2009Jun,Liu2010May} and bilayer graphene~\cite{Volk2011Aug} DQDs on SiO$_2$ where a highly non-monotonous gate voltage dependence has been observed, which has been attributed to disorder and a varying  potential landscape surrounding the QDs.
Beside the capacitive coupling, the interdot tunnel coupling $t_\mathrm{m}$ increases with the electron occupation as can be observed in the charge stability diagram (Fig.~\ref{f3}(a)). An increasing gate voltage weakens the confinement leading to a larger overlap of the electron wave functions. At even higher gate voltages the tunnel barrier becomes fully transparent leaving a single QD. 

Figs.~\ref{f3}(b) and ~\ref{f3}(d) show low ($V_\mathrm{SD} = 0.1$~mV) and high ($V_\mathrm{SD} = 1$~mV) bias charge stability diagrams of the single electron transition $(0,1) \rightarrow (1,0)$. 
To determine the tunnel coupling in this regime, we measure the current as a function of the detuning energy $E_\mathrm{det}$ at a finite bias voltage $V_\mathrm{SD} =1$~mV (see Fig.~\ref{f3}(e)).
As we assume an electron temperature on the order of 100~mK $\approx 8.6~\mu$eV, the condition $eV_\mathrm{SD} >\!\!> k_\mathrm{B}T$ is fulfilled and thus the line width of the resonance at zero detuning is temperature independent. We fit the current data according to a model assuming a Lorentzian line shape leading in the limit $t_\mathrm{m} <\!\!< \Gamma_\mathrm{L,R}$ ~\cite{Fringes2012Jan,Liu2010May,Wiel2002Dec,Stoof1996Jan} to
\begin{equation*}
    I(E_\mathrm{det}) = \frac{4et_m^2 / \Gamma_R}{1+(2E_\mathrm{det}/h\Gamma_R)^2},
\end{equation*}
where $\Gamma_\mathrm{L,R}$ are the tunnel rates to the left and right lead, respectively. The fit yields (i) an interdot tunnel coupling of $h t_\mathrm{m} = 8.5~\mu\mathrm{eV}$ corresponding to an interdot tunnel rate of 2.1~GHz and (ii) a dot-lead tunnel coupling of $h \Gamma_\mathrm{R} = 250~\mu\mathrm{eV}$.

\begin{figure}[]
	\centering
\includegraphics[draft=false,keepaspectratio=true,clip,width=0.95\linewidth]{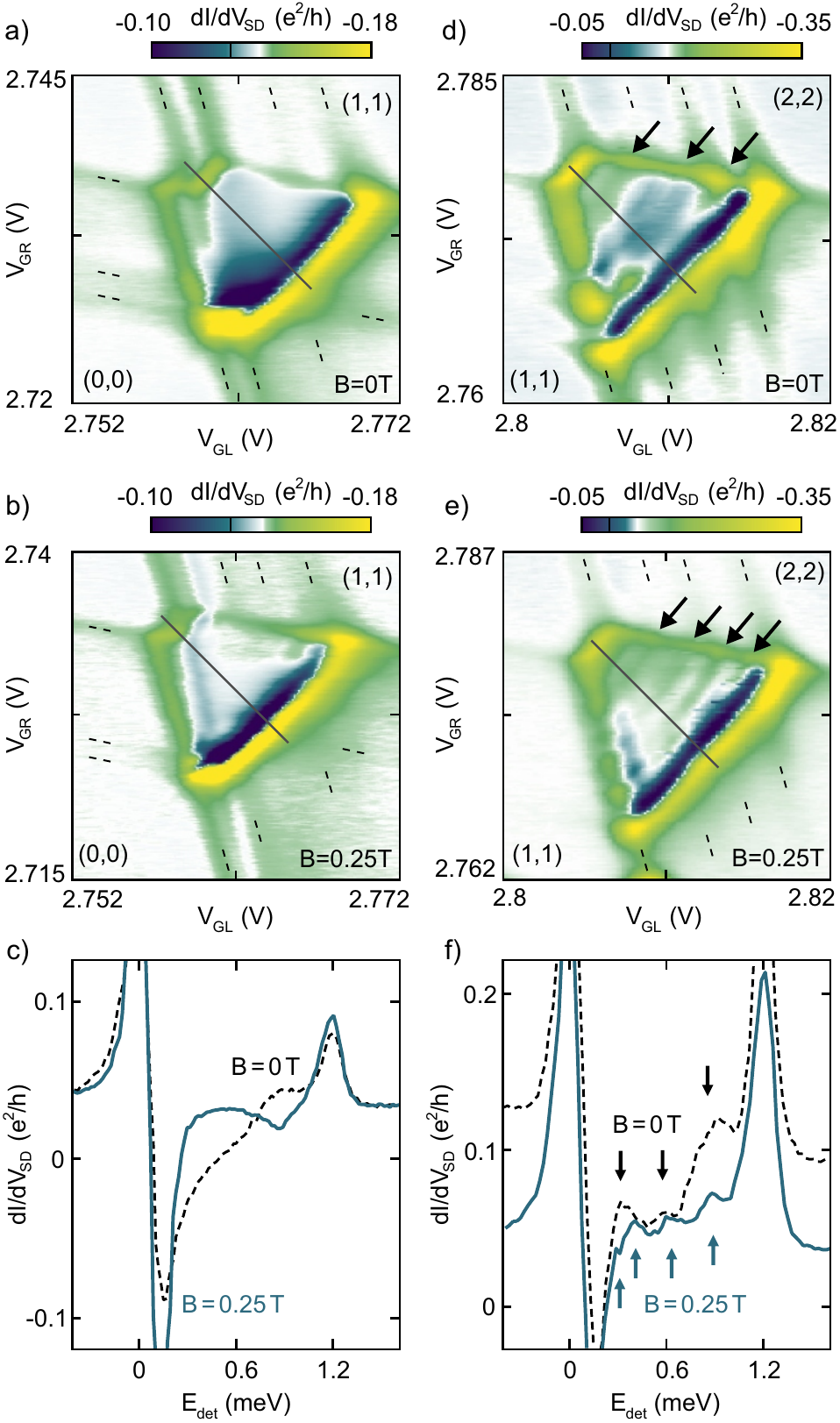}
\caption[Fig04]{Charge stability diagrams showing individual triple points of the DQD measured at zero magentic field and finite out-of-plane magnetic field as indicated in each panel ($V_\mathrm{SD} = 1.2$~mV).
\textbf{(a, b)} Triple point at the addition of the first electrons (transition $(0,0) \rightarrow (1,1)$)
\textbf{(c)} Current through the DQD measured along the interdot detuning axis of the $(0,0) \rightarrow (1,1)$ transition (see solid gray line in panels (a) and (b)). 
\textbf{(d, e)} Triple point at the transition $(1,1) \rightarrow (2,2)$. Arrows indicate excited states, dashed lines emphasize co-tunneling lines.
\textbf{(f)} Current through the DQD measured along the interdot detuning axis of the $(1,1) \rightarrow (2,2)$ transition (see solid gray line in panels (d) and (e)). Arrows highlight excited state transitions. 
} 
\label{f4}
\end{figure}

In Fig. 4(a-d) we show finite bias charge stability diagrams for two different triple points measured at zero and at an out-of-plane magnetic field of $B = 0.25$~T. Interestingly, the triple point of the $(0,0) \rightarrow (1,1)$ transition is almost unaffected by the magnetic field (compare Figs.~\ref{f4}(a) and \ref{f4}(b)).
An excited state is visible as a co-tunneling line parallel to the left edge inside the bias triangle. However, excited states parallel to the base line of zero detuning are absent which can also be seen in a cut along the detuning axis (see Fig.~\ref{f4}(c)). This can be explained in the single particle picture by interdot tunneling processes conserving the spin and valley degree of freedom, allowing only aligned ground state transport. A set of faint co-tunneling lines outside the bias triangle (emphasized by dashed lines in Figs.~\ref{f4}(a) and 4(b)) indicates the presence of a spectrum of excited states which become apparent due to strong tunnel coupling of the DQD to the reservoirs. They become accessible due to inelastic co-tunneling processes if an excited state of one of the QDs is in resonance with the Fermi level of the neighboring reservoir. 

In contrast, the triple point at the $(1,1) \rightarrow (2,2)$ transition shows three excited states within the bias triangle (see arrows in Fig.~\ref{f4}(d)). Their energies measure approximately 0.3, 0.64 and 0.98~meV. Outside the bias triangle, a set co-tunneling lines matches their energies. In contrast to the afore described case, we probe interdot transitions between the (2,1) and (1,2) state, i.e. none of the QDs is empty.
Thus, the description in the single particle picture is no longer valid and interaction effects such as exchange are becoming relevant~\cite{Kurzmann2019Jul}, leading to the observation of an enriched spectrum of available transitions. 
Under the influence of an out-of-plane field ($B = 0.25$~T, see Fig.~\ref{f4}(e)), four excited states at energies of 0.25, 0.46, 0.69 and 0.94 meV can be observed in the bias window (see also Fig.~\ref{f4}(f)). 

To conclude, we studied a bilayer graphene double quantum dot device in the few-electron regime. Finger gates are used to modulate the band profile along a 1D channel defined by metallic split gates.
The device shows a high uniformity which can be seen from finite bias spectroscopy data of two QDs formed independently. By two gates we form a DQD and enable the control of the number of charge carriers on each of the QDs from the few-electron regime down to the very last electron. The interdot tunnel coupling is affected by the same gate voltages such that a transition into a single quantum dot is observed once the total occupation of the DQD exceeds about eight electrons. At a finite bias voltage, we can resolve the excited state spectrum of the DQD. The absence of excited state transitions in the first bias triangle is in agreement with spin and valley conserving interdot tunneling processes. 
Our measurements also show limitations in the current device design and point towards further improvements:
A third gate layer allowing to implement interdigitated finger gates may offer the possibility to gain individual control over the electrochemical potentials and the tunnel barriers. A similar gate architecture has been demonstrated in Si and Ge QD arrays~\cite{Lawrie2019Sep,Mills2019Mar}.
However, with the precise control of the number of electrons in a DQD, we nevertheless meet an important requirement for making such a device a suitable building block for a spin qubit. The measured interdot tunnel coupling on the order of 2~GHz is in a regime compatible with state-of-the-art spin qubit devices~\cite{Huang2019Apr}. 
\newline
\newline
\textbf{Aknowledgements} The authors thank S.~Trellenkamp, F.~Lentz and D.~Neumaier for their support in device fabrication.
This project has received funding from the European Union's Horizon 2020 research and innovation programme under grant agreement No. 785219 (Graphene Flagship) and from the European Research Council (ERC) under grant agreement No. 820254, the Deutsche Forschungsgemeinschaft (DFG, German Research Foundation) under Germany's Excellence Strategy - Cluster of Excellence Matter and Light for Quantum Computing (ML4Q) EXC 2004/1 - 390534769, through DFG (STA 1146/11-1), and by the Helmholtz Nano Facility~\cite{Albrecht2017May}. Growth of hexagonal boron nitride crystals was supported by the Elemental Strategy Initiative conducted by the MEXT, Japan and the CREST(JPMJCR15F3), JST.


\end{document}